\documentclass[prl, twocolumn, preprintnumbers, superscriptaddress]{revtex4}%

\usepackage{amssymb,hyperref, amsmath}
\usepackage[dvips]{color}
\usepackage[dvips]{graphicx}
\usepackage{epsfig}

\begin{document}

\bibliographystyle{apsrev}

\preprint{JLAB-THY-09-1061}

\title{Highly excited and exotic meson spectrum from dynamical lattice QCD}

\author{Jozef J. Dudek}
\email{dudek@jlab.org}
\affiliation{Jefferson Laboratory, 12000 Jefferson Avenue,  Newport News, VA 23606, USA}
\affiliation{Department of Physics, Old Dominion University, Norfolk, VA 23529, USA}

\author{Robert G. Edwards}
\affiliation{Jefferson Laboratory, 12000 Jefferson Avenue,  Newport News, VA 23606, USA}

\author{Michael J. Peardon}
\affiliation{School of Mathematics, Trinity College, Dublin 2, Ireland}

\author{David G. Richards}
\affiliation{Jefferson Laboratory, 12000 Jefferson Avenue,  Newport News, VA 23606, USA}

\author{Christopher E. Thomas}
\affiliation{Jefferson Laboratory, 12000 Jefferson Avenue,  Newport News, VA 23606, USA}

\collaboration{for the Hadron Spectrum Collaboration}

\begin{abstract}
Using a new quark-field construction algorithm and a large variational
basis of operators, we extract a highly excited isovector meson
spectrum on dynamical anisotropic lattices.  We show how carefully
constructed operators can be used to reliably identify the continuum
spin of extracted states, overcoming the reduced cubic symmetry of the lattice.  Using this method we extract, with
confidence, excited states, states with exotic quantum numbers
($0^{+-}$, $1^{-+}$ and $2^{+-}$) and states of high spin, including
for the first time in lattice QCD, spin-four states.
\end{abstract}


\maketitle 

\paragraph{Introduction:}
The spectroscopy of excited meson states is enjoying a renaissance
through the observations of multiple new states in the charmonium
sector. This will continue through the forthcoming experimental efforts at GlueX, BES III and
PANDA that will probe the spectroscopy of mesons in both the light and
charm sectors. New states demand explanation within QCD and may offer insight
into the appropriate degrees-of-freedom of low energy QCD. A
particular example is mesons of exotic $J^{PC}$, those states whose
quantum numbers cannot be constructed from a quark-antiquark bound
state, and whose existence may signal the influence of explicit
gluonic degrees of freedom.

Lattice QCD provides an {\em ab initio} method for the determination
of the hadron spectrum.  This approach to spectroscopy necessitates
methods for measuring the two-point correlation functions of field
operators with the selected quantum numbers under investigation.
However, it has proven difficult to extract precise information from
lattice QCD about excited states, states of high spin and states with
exotic $J^{PC}$. In this letter we will present results using a large
basis of composite QCD operators and a variational analysis method
which show that such extractions are now possible.

Access to states of spin-two or higher requires operators with
spatially separated quark fields. To facilitate this kind of
construction, a new quark-field construction algorithm, called
``distillation'', was developed~\cite{Peardon:2009gh} recently which
enables efficient calculations of a broad range of hadron correlation
functions, including those with spatially separated quark fields.

In Euclidean space, excited-state contributions to correlation functions
decay faster than the ground-state and at large times are swamped by
the larger signals of lower states. In improving our ability to
extract excited states, better temporal resolution of correlation
functions proves extremely helpful. An anisotropic lattice, where the
temporal direction is discretized with a finer grid spacing than its
spatial counterparts, is one means to provide this resolution while
avoiding the increase in computational cost that would come from
reducing the spacing in all directions. To this end, a large-scale
program has been initiated to generate dynamical anisotropic gauge fields with
two light clover quarks and one strange quark~\cite{Edwards:2008ja,Lin:2008pr}. 

In this work, these anisotropic lattices are combined with the
distillation technique for the construction of quark-antiquark
operators with multiple derivative insertions. Only the connected
Wick contractions are computed, giving access to isovector quantum
numbers. For this first investigation, the three-flavor
degenerate-quark-mass dataset is used ($m_\pi =m_K=m_\eta \approx 800 \,\mathrm{MeV}$), with lattice spacings $a_s \sim 0.12\,\mathrm{fm}, \,
a_t^{-1} \sim 5.6\, \mathrm{GeV}$ and a spatial lattice
extent of $\sim 2\,\mathrm{fm}$.  We will argue later that using a
relatively large quark mass in this first study reduces complications
due to mesons being able to decay into multi-meson states.

\paragraph{Spin on a cubic lattice:}
Lattice QCD computations consider the theory discretised on a
four-dimensional Euclidean hypercubic grid.  The reduced
three-dimensional rotational symmetry with respect to the continuum
introduces complications when one wishes to study particles of a
particular spin, since spin no longer identifies irreducible
representations of the cubic symmetry
group\cite{Johnson:1982yq}. There are five single-cover lattice
\emph{irreps} for each parity and charge-conjugation: $A_1, T_1, T_2,
E, A_2$.
The distribution of the various $M$ components of a spin-$J$ meson
into the lattice irreps is known as \emph{subduction}, the result of
which is displayed in Table \ref{subduce}. In the continuum limit, the
full rotational symmetry is restored and the components subduced into
different irreps will be degenerate, whereas at finite lattice spacing
they will be split by an amount scaling with at least one
power of the
lattice spacing, $a_s$.

\begin{table}
\begin{tabular}{ccl}
$J$ & & irreps \\
\hline
$0$ & & $A_1(1)$ \\
$1$ & & $T_1(3)$ \\
$2$ & & $T_2(3) \oplus E(2)$\\
$3$ & & $T_1(3) \oplus T_2(3) \oplus A_2(1)$\\
$4$ & & $A_1(1) \oplus T_1(3) \oplus T_2(3) \oplus E(2)$
\end{tabular}  
\caption{Continuum spins subduced into lattice irreps $\Lambda(\mathrm{dim})$.}
\label{subduce}
\end{table}

\begin{figure}
 \centering
\psfig{width=0.48\textwidth,file=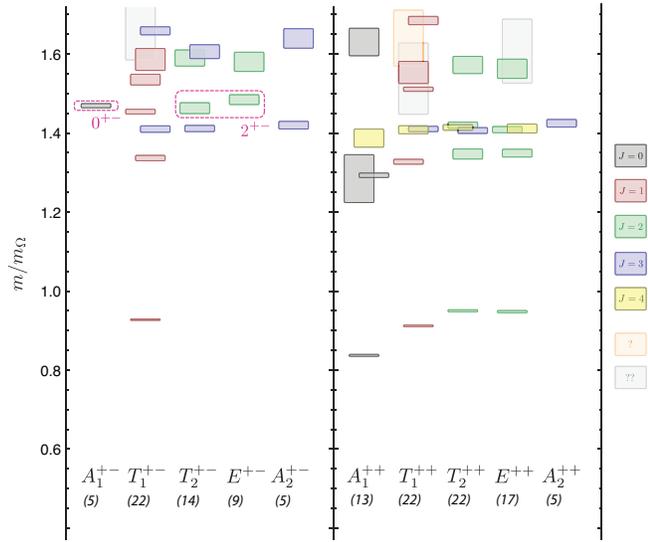}  
 \caption{Extracted spectrum of states in the $PC=+-,++$ channels
   displayed by lattice irrep.  The number of operators in each irrep
   is given below the irrep label. All masses scaled by the $\Omega$ baryon mass as extracted on this lattice \protect \cite{Lin:2008pr}. Boxes represent the extracted mass and
   one sigma statistical uncertainties. Color coding indicates
   continuum spin identification. Orange boxes have well determined
   masses but undetermined spin. Grey boxes have masses that are not
   well determined by the variational fitting method. States with exotic
   quantum numbers $0^{+-}$ and $2^{+-}$ are highlighted.}
 \label{spectrum1}
\end{figure}

\begin{figure}
 \centering
\psfig{width=0.48\textwidth,file=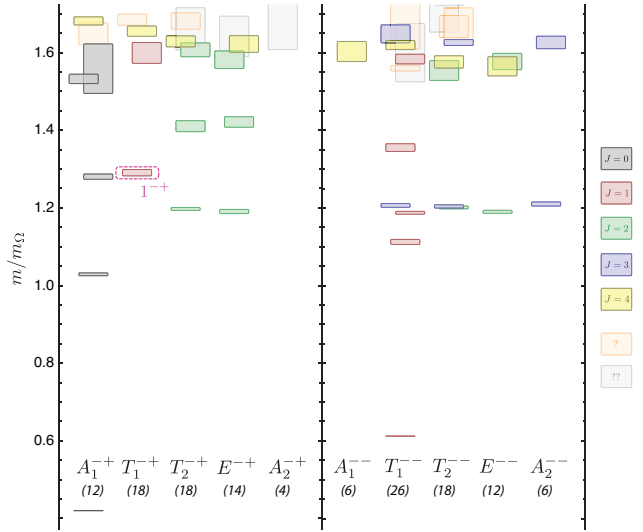}        
\caption{As previous but for $PC=-+,--$. The lowest lying exotic $1^{-+}$ is
   highlighted.}
 \label{spectrum2}
\end{figure}

This suggests a simple method to assign continuum spins by attempting
to identify degeneracies across lattice irreps compatible with the
subduction patterns in Table \ref{subduce}. Unfortunately the
empirical meson spectrum shows a number of approximate degeneracies
that may be confused with those originating through discretisation. As
an example consider the $\chi_{c0,1,2}$ states in charmonium, split
only by a small spin-orbit force. These states would appear in a
lattice computation as a single state in each of $A_1^{++}, T_1^{++}, T_2^{++} $ and
$E^{++}$ and could easily be mistaken with a single $J=4$ state split by
discretisation effects. In the high lying part of the calculated
spectrum, shown in Figures \ref{spectrum1} and \ref{spectrum2}, we
observe considerable degeneracy that renders spin-identification by
this method virtually impossible.
In this letter we consider using the additional information embedded in 
the overlaps of states on to carefully constructed operators at zero momentum.

\paragraph{Meson operators:}
By using a circular basis for both spatial derivatives and the three-vector-like gamma
matrices ($\gamma_i$), we can utilise the SO(3) Clebsch-Gordan
coefficients to construct continuum operators of definite spin. For
example, with one derivative and a vector gamma matrix we can construct
operators of overall spin $J=0,1,2$:
\begin{equation}
 (\Gamma \times D^{[1]}_{J_D=1} )^{J, M} \equiv \big\langle 1, m_1 ; 1, m_2 \big| J, M
  \big\rangle\;  \widetilde{\bar{\psi}}\, \Gamma_{m_1}
  \overleftrightarrow{D}_{m_2} \widetilde{\psi} \nonumber
\end{equation}
where repeated $m$ indices are summed. In the distillation framework,
the fermion fields, $\widetilde{\psi}$ are smeared using a low-rank 
filtering operator. 

 In the case of two derivatives we couple into a definite spin before 
coupling to the gamma matrix:
\begin{multline}
(\Gamma \times D^{[2]}_{J_D})^{J,M} \equiv \big\langle 1, m_3 ; J_D, m_D \big| J, M
  \big\rangle \\ \quad \quad \quad \times \big\langle 1, m_1 ; 1, m_2 \big| J_D, m_D
  \big\rangle \times \widetilde{\bar{\psi}}\, \Gamma_{m_3}
  \overleftrightarrow{D}_{m_1} \overleftrightarrow{D}_{m_2} \widetilde{\psi}. \nonumber
\end{multline}
For three derivatives combining the outermost derivatives together
first ensures definite charge-conjugation:
\begin{multline}
  (\Gamma \times D^{[3]}_{J_{13}, J_D})^{J,M} \equiv \big\langle 1, m_4 ; J_D, m_D \big| J, M
  \big\rangle \\ \times \big\langle 1, m_2 ; J_{13}, m_{13} \big| J_D, m_D
  \big\rangle \times \big\langle 1, m_1 ; 1, m_3 \big| J_{13}, m_{13}
  \big\rangle \\ \times \widetilde{\bar{\psi}} \,\Gamma_{m_4}
  \overleftrightarrow{D}_{m_1} \overleftrightarrow{D}_{m_2}
  \overleftrightarrow{D}_{m_3} \widetilde{\psi}. \nonumber
\end{multline}
This scheme can be extended to any desired number of
covariant derivatives, which in practical computations are replaced by
gauge-covariant finite differences. The gauge links appearing in these
differences are stout-smeared to reduce UV fluctuations. To be of any real use in lattice
calculations these operators of definite continuum spin, $J$, must be
\emph{subduced} into the irreducible representations of the cubic
lattice rotation group ($\Lambda = \{ A_1, T_1, T_2, E, A_2
\}$). Noting that each class of operator is closed under rotations,
the subductions can be performed using known linear combinations of
the $M$ components for each $J$:
\begin{eqnarray*}
\lefteqn{{\cal O}^{[J]}_{\Lambda,\lambda} \equiv (\Gamma \times D^{[n_D]}_{\ldots})^J_{\Lambda, \lambda} = }\\
& & \sum_M {\cal
     S}^{J,M}_{\Lambda, \lambda}   \; (\Gamma \times
   D^{[n_D]}_{\ldots})^{J,M} \equiv \sum_M S^{J,M}_{\Lambda,\lambda} {\cal O}^{J,M}
\end{eqnarray*}
where $\lambda$ is the ``row'' of the irrep. Note that, although ${\cal
  O}^{[J]}_{\Lambda, \lambda}$ can have an overlap with all spins contained
within $\Lambda$, it still carries the memory of the $J$ from which
it was subduced, a feature we exploit below.

\paragraph{Spectral analysis:}
For each lattice irrep $\Lambda^{PC}$ the full matrix of correlators
$C_{ij}(t)$ was computed with equivalent rows ($\lambda$) averaged over.  The 
dimension of the matrix is therefore equal to the number of operators 
constructed in that irrep.

The correlation matrix can be described by a spectral decomposition $C_{ij}(t) = \sum_{\mathfrak{n}} \frac{Z^{\mathfrak{n}*}_i
  Z^{\mathfrak{n}}_j }{2 m_{\mathfrak{n}} } e^{- m_{\mathfrak{n}} t}$ (we only consider zero momentum),
where $Z^{\mathfrak{n}}_i = \langle 0 | {\cal O}_i | \mathfrak{n}
\rangle$ encodes the overlap of state $\mathfrak{n}$ on to operator
${\cal O}_i$. An optimal method (in the variational sense
\cite{Luscher:1990ck,Blossier:2009kd}) to extract mass and $Z$ information from the
matrix of correlators is by solution of a generalised eigenvalue
problem, $C_{ij}(t) v^{\mathfrak{n}}_j = \lambda^{\mathfrak{n}}(t,t_0) C_{ij}(t_0)
v^{\mathfrak{n}}_j$ where the eigenvectors $v^{\mathfrak{n}}$ are
  related to the $Z$ by $Z^{\mathfrak{n}}_j = \sqrt{2m_{\mathfrak{n}} }
e^{m_{\mathfrak{n}}  t_0/2} v^{\mathfrak{n}*}_i C_{ij}(t_0)$. 
Our implementation of this approach is described in
\cite{Dudek:2007wv}.

The extracted spectrum across lattice irreps, including all operators
with up to three derivatives, is shown in Figures \ref{spectrum1} and
\ref{spectrum2}.  We give the number of operators in each irrep and 
the color coding indicates continuum spin assignment 
suggested by a method we now describe. 

Our particular choice of operator construction offers us a method to
identify the continuum spin of a state.  We take advantage of the fact
that, at the lattice spacing we work, we expect 
lattice operators acting on extended objects such as mesons to 
behave in a manner 
reasonably close to the full rotational 
symmetry.  In the continuum our operators are of definite spin such that 
$\langle 0 | {\cal O}^{J,M} |J', M'\rangle = Z^{[J]} \delta_{J,J'} \delta_{M,M'}$ 
and so $\langle 0  | {\cal O}^{[J]}_{\Lambda, \lambda} | J', M\rangle = {\cal
  S}^{J,M}_{\Lambda, \lambda} Z^{[J]} \delta_{J,J'}$. 
$Z^{[J]}$ is a single number of dynamical origin describing the overlap 
of the state of spin $J$ on to the operator used. We form a correlator 
in a given irrep $\Lambda$ and average over equivalent rows, $\lambda$, 
\begin{equation}
  \tfrac{1}{\mathrm{dim}(\Lambda)} \sum_\lambda
  C^{[\Lambda]}_{\lambda \lambda} \equiv  \tfrac{1}{\mathrm{dim}(\Lambda)} \sum_\lambda
  \langle 0 | {\cal O}^{[J]}_{\Lambda, \lambda}  {\cal
    O}^{[J]\dag}_{\Lambda, \lambda} | 0 \rangle.  \nonumber
\end{equation}
Inserting a complete set of meson states between the operators and using the fact 
that the subduction coefficients form an orthonormal matrix, 
$\sum_M {\cal S}^{J,M}_{\Lambda, \lambda} {\cal
  S}^{J,M*}_{\Lambda', \lambda'} = \delta_{\Lambda, \Lambda'}
\delta_{\lambda, \lambda'}$, 
we obtain terms proportional to $Z^{[J]*} Z^{[J]}$; these terms do not depend upon
which $\Lambda$ we have subduced into. Hence for example a $J=3$ meson
created by a $[J=3]$ operator will have the same $Z$ value in each of
the $A_2, T_1, T_2$ irreps. Since this derivation uses smoothed, semi-classical
fields it is valid in the continuum limit and at finite lattice spacing we 
expect there to be small deviations from 
equality due to discretisation effects.

\begin{figure}
 \centering
\psfig{width=0.45\textwidth,file=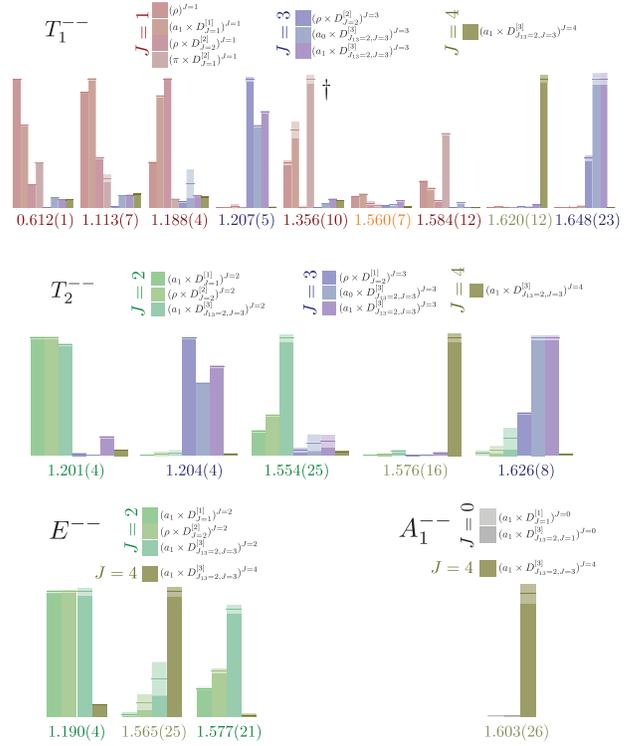}  
 \caption{Overlaps, $Z$, of a selection of operators on to states in each lattice
   irrep, $\Lambda^{--}$. $Z$'s are normalised so that the largest value across all
   states is equal to $1$. Lighter area at the head of each bar
   represents the one sigma statistical uncertainly.}
 \label{histo}
\end{figure}

\begin{figure*}
 \centering
\psfig{width=0.87\textwidth,file=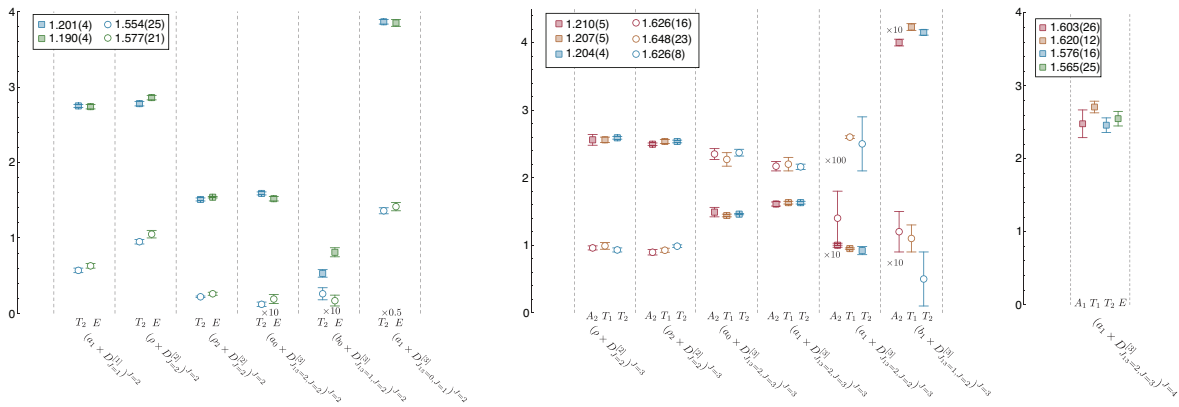}  
 \caption{A selection of $Z$ values across irreps $\Lambda^{--}$ for states suspected of being
   $J=2,3,4$ respectively. }
 \label{Zs}
\end{figure*}

We take advantage of these properties to identify the spin of the
extracted states in the following way. Firstly, we consider the
relative magnitudes of the extracted $Z$ values for various states. 
Figure \ref{histo} shows that for the $J^{--}$ mesons, each state has
large overlap only onto operators of a single spin. The second stage
of the identification requires us to match states in different irreps
and compare their $Z$ values with common operators subduced across
irreps. 
As shown in Figure \ref{Zs}, these values agree well. Any
small discrepancy could be attributed to two causes: discretisation errors from the use of simple
central-difference operators to represent derivatives or the effect of renormalisation. These operators act on 
smoothed gluonic and quarks fields and this 
eliminates fluctuations at the cut-off scale so the latter effects will most likely be very small.

Using this method we have extracted a large number of states with all
possible $PC$ combinations and confidently identified the spin of
these states; the spectrum is shown in Figures \ref{spectrum1} and
\ref{spectrum2}.  As well as extracting many excited states, we have
for the first time identified states with spin four: $4^{++}$,
$4^{-+}$ and $4^{--}$.  We have extracted states with exotic quantum
numbers ($0^{+-}$, $1^{-+}$ and $2^{+-}$) and these are highlighted in
the figures.  The presence of these exotics likely points to the influence of
explicit gluonic degrees of freedom.

Figure \ref{histo} shows that the third excited vector state ($m/m_\Omega \sim
1.35$, marked with $\dag$) has a qualitatively different pattern of $Z$ values
compared to the lighter vectors.  Notably, there is large
overlap with the $(\pi \times D^{[2]}_{J=1})^{J=1}$ operator.  $D^{[2]}_{J=1}$ corresponds to the commutator of two covariant
derivatives which vanishes in the absence of a gluonic field. 
This commutator is proportional to the
field strength tensor and so the significant overlap hints at a
gluonic component.  This suggests an identification of this state as
a crypto-exotic vector hybrid, although the non-zero overlap onto
$\widetilde{\bar{\psi}} \gamma_i \widetilde{\psi}$ suggests some mixing with a
conventional vector state.


\paragraph{Two-meson states:}
We might expect to observe an abundance of two-meson states above
$2m_\pi \sim 0.85 m_\Omega$, but such states are not apparent in
our extracted spectrum. This is most clearly seen in the $A_1^{--}$
channel where the lightest state extracted is a $J=4$ state above $1.5 m_\Omega$, while a pseudoscalar-vector state with the minimum relative
momentum allowed in our finite box would be expected close to $1.2 m_\Omega$. The operators used in this study featured only a single $\psi,
\bar{\psi}$ field pair and so do not have overlap onto quark Fock
states higher than $q\bar{q}$. QCD dynamics can act to mix $q\bar{q}$
Fock states with two-meson basis states to form mesonic eigenstates. This
mixing is expected to be significant when a discrete lattice two-meson
state is degenerate with a ``single meson'' to within that meson's
continuum decay width. At this relatively heavy quark mass, we expect
low-lying resonances to have small widths due to reduced phase-space
for their decay and hence for there to be only small mixing with
two-meson states, perhaps explaining our lack of observation of such
states. A calculation similar to the one reported herein has been carried out on a lattice of spatial extent $\sim 2.4 \,\mathrm{fm}$. The extracted spectrum is found to be identical within statistical fluctuations to that presented here. This is more evidence that we are not seeing two-meson states since their allowed relative momentum, and hence their energy levels, would have changed significantly.
These issues can be properly investigated by including in the
variational basis operators featuring a product of two fermion
bilinears, expected to have good overlap on to two-meson states. This
work is underway.


\paragraph{Summary:}
We have demonstrated a lattice QCD operator construction that enables the
identification of continuum spin with some confidence. Using
distillation technology to construct the correlators, and a
variational analysis to study them, we have extracted an excited state
spectrum featuring well-determined states with exotic quantum numbers
and, for the first time, states of spin-4.

It is notable that our extracted spectrum has both features of the
$n^{2S+1}L_J$ state assignment of bound-state quark models and also
states that do not seem to lie within that classification. We believe
that this study is seeing a full spectrum of QCD mesons which
includes exotic and non-exotic hybrid mesons\cite{Dudek:2008sz}.

\paragraph{Acknowledgments - }
We thank our colleagues within the Hadron Spectrum Collaboration. The Chroma software suite~\cite{Edwards:2004sx} was used to perform this work 
on clusters at Jefferson Laboratory using time awarded under the USQCD 
Initiative. Authored by Jefferson Science Associates, LLC under U.S. DOE Contract No.
DE-AC05-06OR23177. 


\bibliography{NF3_letter}

\end{document}